\documentclass[aps,prb,twocolumn,longbibliography,superscriptaddress]{revtex4-1}
\usepackage{epsfig}
\usepackage{epstopdf}
\usepackage{amsmath}
\usepackage{amsfonts}
\usepackage{amssymb}
\usepackage{hyperref}
\usepackage{bm}
\usepackage{makecell}
\usepackage{rotating}
\usepackage{hyperref}
\usepackage{graphicx}
\usepackage{dcolumn}
\usepackage{bm}
\usepackage{color}

\usepackage{tikz,xcolor,hyperref}
\definecolor{lime}{HTML}{A6CE39}
\DeclareRobustCommand{\orcidicon}{%
	\begin{tikzpicture}
	\draw[lime, fill=lime] (0,0)
	circle [radius=0.16]
	node[white] {{\fontfamily{qag}\selectfont \tiny ID}};
	\draw[white, fill=white] (-0.0625,0.095)
	circle [radius=0.007];
	\end{tikzpicture}
	\hspace{-2mm}
}

\foreach \x in {A, ..., Z}{%
	\expandafter\xdef\csname orcid\x\endcsname{\noexpand\href{https://orcid.org/\csname orcidauthor\x\endcsname}{\noexpand\orcidicon}}
}

\begin{document}

\title{Orbital order and ferromagnetism in LaMnO$_3$ doped with Ga}
\author{C. Autieri\orcidA}\email{autieri@magtop.ifpan.edu.pl}
\affiliation{International Research Centre Magtop, Institute of Physics, Polish Academy of Sciences,
Aleja Lotnik\'ow 32/46, PL-02668 Warsaw, Poland}
\author{M. Cuoco\orcidB}
\affiliation{Consiglio Nazionale delle Ricerche (CNR-SPIN), I-84084 Fisciano (SA), Italy} 
\affiliation{Dipartimento di Fisica ``E. R. Caianiello'', Universit\`a di
Salerno, I-84084 Fisciano (SA), Italy}
\author{G. Cuono\orcidC}\email{gcuono@magtop.ifpan.edu.pl}
\affiliation{International Research Centre Magtop, Institute of Physics, Polish Academy of Sciences,
Aleja Lotnik\'ow 32/46, PL-02668 Warsaw, Poland}
\author{S. Picozzi\orcidD}
\affiliation{Consiglio Nazionale delle Ricerche (CNR-SPIN), Via Vetoio, I-67010 L'Aquila, Italy}
\author{C. Noce\orcidE}
\affiliation{Dipartimento di Fisica ``E. R. Caianiello'', Universit\`a di
Salerno, I-84084 Fisciano (SA), Italy}
\affiliation{Consiglio Nazionale delle Ricerche (CNR-SPIN), I-84084 Fisciano (SA), Italy}

\begin{abstract}
We study from first principles the magnetic, electronic, orbital and structural properties of the LaMnO$_3$ doped with gallium replacing the Mn-site. 
The gallium doping reduces the Jahn-Teller effect, and consequently the bandgap.
Surprisingly, the system does not go towards a metallic phase because of the Mn-bandwidth reduction.
The Ga-doping tends to reduce the orbital order typical of bulk antiferromagnetic LaMnO$_3$ and consequently weakens the antiferromagnetic phase. 
The Ga-doping favors the G-type orbital order and layered-ordered ferromagnetic perovskite at x=0.50, both effects contribute to the formation of the insulating ferromagnetic phase in LaMn$_{1-x}$Ga$_{x}$O$_3$.
\end{abstract}
\pacs{31.15.A-,71.10.Hf}
\maketitle

\section{Introduction}

In recent years the series
La$_{1-x}$A$_{x}$MnO$_3$, where A is a divalent metal, has been the
object of systematic investigations \cite{Cheong00,Offi21,Bhattacharya14}.
This is not only due to the discovery of the giant magnetoresistance in several members of the series but also for the strong interplay among orbital, lattice, spin and charge degrees of freedom exhibited by the compounds of the series. This interplay results in
a large variety of magnetic arrangements and phase transitions depending on the hole doping.
Concerning the giant magnetoresistance phenomena, it is observed in the ferromagnetic metallic phase \cite{Tokura00}. The correlation between ferromagnetism and metallic behaviour has been explained by the double exchange mechanism \cite{Zener51} whereas the electron-phonon coupling is mainly due to the Jahn-Teller (JT) effect \cite{Kanamori60,Pavarini10}. Interestingly, the ferromagnetic phase of La$_{1-x}$A$_{x}$MnO$_3$ has been used to study the interplay between ferromagnetism and ferroelectricity at the oxide interface\cite{Autieri14,Paul14,Hausmann17} and to create high-temperature ferromagnetism in the 2D limit\cite{Piamonteze21,Zakharova21}.\\
Looking at the mother compound LaMnO$_3$, it has been considered as the prototype example of a cooperative JT system and orbital-order state \cite{Fazekasbook,Yamasaki06}. Indeed, LaMnO$_3$ shows an orthorhombic unit cell with a cooperative tetragonal deformation of the MnO$_6$ octahedra, JT like, at room temperature. Moreover, the system is characterized by a large octahedral tilt/rotation pattern of the $a^{-}a^{-}c^{+}$ type\cite{Lufaso01}, following the Glazer notation, leading to the Pnma (or Pbnm in non-standard setting) space group, whose structural parameters are shown in Table \ref{UndopedCompound}.
\begin{table}[!ht]
\caption[{[tabhop}]{Structural parameters of the $Pbnm$ (No. 62 in the International Tables) structure of the parent compound LaMnO$_3$ as reported by Elemans {\it et al.} \cite{Elemans71}, $a=5.532$ {\AA}, $b=5.742$ {\AA}, $c=7.668$ {\AA} at 4.2 K.
}

\begin{center}\label{UndopedCompound}
\begin{tabular}{l|l|l|l}
\hline
\hline
Atoms \&  Wyckoff site    & x & y & z \\
\hline
La (4c)  &  -0.010   &  0.049 &  0.250 \\
Mn (4b)  &    0.500  &  0.000 &  0.000 \\
O(1) (4c)&   0.070  & 0.486  & 0.250 \\
O(2) (8d)&   0.724  & 0.309  &   0.039 \\
\hline
\hline
\end{tabular}
\end{center}
\end{table}
This compound develops long-range antiferromagnetic (AFM) ordering of type A below T$_{N}$=140 K. Specifically, the
manganese moments are aligned in the [010] direction with the spins coupled ferromagnetically in the $ab$ plane and antiferromagnetically along the $c$-axis\cite{Kovaleva04}. An exotic ferromagnetic state was found theoretically in heterostructures based on LaMnO$_3$, showing that the interplay between the band reduction due to the dimensionality with strain and interface effects can lead the LaMnO$_3$ to the ferromagnetic phase.\cite{Banerjee20,Cossu21}
However, this band reduction due to the dimensionality is absent in bulk systems.

Surprisingly, the replacement of the magnetic JT Mn$^{3+}$ ion by an isoelectronic non-magnetic non Jahn-Teller ion such as Ga$^{3+}$ induces long-range ferromagnetism \cite{Goodenough03,Pastorino,Ver02,Blasco,Sanchez06} without carrying hole doping. Without hole doping, the double-exchange mechanism is not active. Nevertheless, the ferromagnetic interactions
can not be ascribed to any of the previous mechanisms.
It was shown indeed the relationship between the
static Jahn-Teller distortion of the MnO$_6$ octahedron and
the orthorhombic distortion of the unit cell in the LaMn$_{1-x}$Ga$_{x}$O$_3$ series \cite{Goodenough03}.
The Mn and Ga ionic radii are 0.785 and 0.760 \AA, respectively \cite{Shannon}, so that the replacement of manganese by the smaller gallium reduces the octahedral distortions and produces the appearance of a spontaneous magnetization\cite{Blasco}.
The distribution of the Ga atoms is uniform due to the missing of the bond formation for d$^{10}$ closed-shell\cite{Nucara14}, while in the limit of strong Ga-concentration the Mn atoms tend to segregate due to the open shells d$^4$ that create bonds at the Fermi level\cite{Aleshkevych15}.
In the limit of strong Ga-concentration, the perovskite systems with Ga$^{3+}$ and Mn$^{3+}$ behave like dilute magnetic semiconductors (DMS)\cite{Aleshkevych15,Jin21} where the d$^4$-d$^4$ magnetic interaction is usually ferromagnetic also in the insulating phase. Different mechanisms were proposed to describe the ferromagnetism in DMS with Mn-d$^4$ electronic configuration\cite{Dietl2000,Litvinov2001,Kaminski2002,Dietl21}, with the superexchange that dominates in absence of band carriers\cite{sliwa2021superexchange,autieri2020momentum}.

The maximum total magnetic moment in LaMn$_{1-x}$Ga$_{x}$O$_3$ is achieved for $x$=0.5.
For x$>$0.6 the lack of Jahn-Teller effect makes the system cubic \cite{Sanchez06}.
Unusually for ferromagnetic compounds, these samples are also electrically insulators.
The gallium doping has dramatic effect at concentrations lower than 5\%, since one Ga atom increases the magnetic moment in an applied magnetic field up to 16 $\mu_B$ per Ga atom \cite{Ver02}.
Investigating similar ferromagnetism in LaMn$_{1-x}$Sc$_x$O$_3$, it was shown that the ordering of the Mn$^{3+}$ Jahn-Teller distortion is not disrupted in the ab-plane for any Sc concentration. This contrasts with the results of the LaMn$_{1-x}$Ga$_{x}$O$_3$, where a regular MnO$_6$ is found for x$>$0.5. Therefore, both LaMn$_{0.5}$Sc$_{0.5}$O$_3$ and LaMn$_{0.5}$Ga$_{0.5}$O$_3$ show a similar ferromagnetic behavior independently of the presence or not of the Jahn-Teller distorted Mn$^{3+}$, pointing to the Mn-sublattice dilution as the main effect in driving ferromagnetism over the local structural effects\cite{Subias16}.

Here, using first principles calculations, we study the electronic, magnetic and orbital properties of the LaMn$_{1-x}$Ga$_x$O$_3$ focusing on the interplay between the Jahn-Teller and the octahedral distortions in the antiferromagnetic phase.
The paper is organized as follows: we present the computational details for the
{\it ab-initio} calculations in Section II, while in Section III we show the results from first principles studies, focusing on the octahedral distortions, the density of state (DOS) and the orbital order. Finally, in Section IV we propose the possible origin of the experimentally detected ferromagnetic phase.

\section{Computational details}

We perform spin-polarized first-principles density functional theory (DFT) calculations
\cite{Kohn64} using the Quantum Espresso program package \cite{QE}, the
GGA exchange-correlation functional of Perdew, Burke, and
Ernzerhof \cite{PBE}, and the Vanderbilt ultrasoft pseudopotentials \cite{Ultrasoft} in
which the La$(5s,5p)$ and Mn$(3s,3p)$ semicore states are included in the valence.
We used a plane-wave energy cut-off of 35 Ry and a Gaussian broadening of 0.01 Ry
as in  the reference \onlinecite{Kova10}.
These values for the plane-wave cutoff and the Gaussian broadening
are used in all calculations presented in this chapter. A
$10\times10\times10$ k-point grid is used in all DOS calculations, while a $8\times8\times8$ grid is used for the relaxation of the internal degrees of freedom.
We optimized the internal degrees of freedom by minimizing the total energy to be less than 10$^{-4}$ Hartree and the remaining forces are smaller than 10$^{-3}$ Hartree/Bohr, while fixing the lattice parameter $a$, $b$ and $c$ to the experimental values \cite{Elemans71,Blasco}.
After obtaining the DFT Bloch bands within GGA, we use the occupation matrix to obtain the orbital order parameter.
For the DOS calculations, we used a Gaussian broadening of 0.02~eV to
have an accurate measurement of the bandgap.

To go beyond GGA, it has been proposed to include the Coulomb repulsion $U$ into the GGA theory giving rise to the so-called GGA+$U$ theory.
GGA+$U$ was first introduced by Anisimov and his coworkers \cite{Anisimov91,Anisimov93}. Here, we use the rotational invariant form introduced by Lichtenstein\cite{Anisimov95} in its spherically averaged and simplified approach\cite{Dudarev98} where there is just an adjustable parameter $U_{eff}=U-J$.
A self-consistent method for the determination of $U_{eff}$ was proposed by Cococcioni et al. \cite{Cococcioni05} Starting from the observation of the non piecewise behaviour of the energy as a function of the occupation number \cite{Anisimov93}, they
implemented a method to take into account the electron screening in the Hubbard repulsion. We used the refined approach suggested by Cococcioni seeking internal consistency for the value of $U_{eff}$.
Once calculated the first value of $U_{eff}$ within the GGA scheme, we performed the Cococcioni technique for the functional GGA+$U_{eff}$ obtaining a correction to the $U_{eff}$ value and repeating the procedure until the correction for the final value of $U_{eff}$ vanishes. We constructed several supercells required by the Cococcioni method and we calculated $U_{eff}$ from the AFM configuration using experimental volume and atomic positions \cite{Pickett96}.
The well converged value for the AFM configuration of LaMnO$_3$ is $U_{eff}$=6.3~eV.
The $U_{eff}$ obtained in the AFM phase is in good agreement with $U_{eff}$ values  between 3.25 and 6.25~eV used for LaMnO$_3$\cite{Anisimov10,Pavarini10} and other Mn-based insulating systems\cite{Keshavarz17,Ivanov16,Ivanov17,Pournaghavi21}.
We use the value $U$=6.3~eV in all the configurations and at all the doping concentrations of LaMn$_{1-x}$Ga$_{x}$O$_{3}$.

The optimization  of manganites within DFT is a very delicate issue and no clear methodology (GGA, LSDA, GGA+U, etc) has been assessed in the literature, each having advantages and disadvantages, as for the description of tilting, JT distortion, lattice parameters etc.
We perform the relaxation of the atomic positions at fixed experimental volume for some doping concentrations in GGA in the AFM spin configuration using several supercells. After the relaxation, we calculate the density of state, the distortion and the orbital order in GGA+$U$ using the $U$ calculated by the Cococcioni method.

Here, we study the properties of the LaMn$_{1-x}$Ga$_{x}$O$_3$ at $x$=0.000, 0.125, 0.250, 0.500 considering a $2\times2\times2$ supercell for x=0.125 and 0.250 while we consider a $\sqrt{2}\times\sqrt{2}\times2$ supercell for the other cases.
We substitute the Ga atom to the Mn atom in the centre of the octahedron. We use the experimental volume from \onlinecite{Elemans71,Blasco} to construct the supercells. The net magnetic moment of the supercell will be different from zero at $x$=0.125, 0.250 due to the presence of an odd number of Mn atoms.
The impurities distribution could be treated with more advances techniques to get accurate quantitative results, however, the results obtained within these approximations return satisfactory qualitative results.

\section{Electronic and structural properties}

Pickett et al. \cite{Pickett96} showed that, in GGA approximation, the gap of AFM phase of LaMnO$_3$ is entirely due to the octahedral distortion. We find the same result also in GGA+$U$, so, we can conclude that the gap in GGA+$U$ is a rough sum of the contribution due to the distortion plus the contribution due to the Coulomb repulsion $U$.

Here, we will examine the influence of Ga-doping of LaMnO$_3$ on the AFM phase. In this case, the Mn$^{3+}$ is in the high-spin state configuration $t_{2g}^{3} e_g^{1}$, and for every Mn atom present in the supercell, there is one occupied $e_g$ level with majority spin. The undoped LaMnO$_3$ exhibits an experimental gap equal to 1.2 eV \cite{Tobe01}, therefore, we expect that the ferromagnetic configuration would be an insulator. Instead, it has been found that ferromagnetic LaMnO$_3$ is a half-metal\cite{Kova10}, and as well know in the literature \cite{Sawada97,Terakura10}, it is very hard to computationally reproduce the ferromagnetic insulating phase in bulk systems since the half-metal phase is very stable against the bandwidth reduction\cite{Cossu14}. For this reason, we will concentrate on the AFM Ga-doped phase of LaMnO$_3$, investigating the octahedral distortions, the density of states and the orbital order, whereas in the next Section we will address the magnetic properties of the system.

\subsection{Octahedral Distortion}

We calculate the evolution of the geometrical properties of the system: the Mn-O-Mn bond angles, the Ga-O-Mn bond angles and other octahedral parameters as the distances between the oxygens and the transition metals.
The geometric bond angles as a function of the Ga-doping are shown in Table \ref{GaOMnangle}. For the undoped case, we found a Mn-O-Mn bond angle of 160$^\circ$, then the bond angles tend to increase slightly when we increase the Ga concentration, though they are not going towards 180$^\circ$ even at x=0.500 when
the Jahn-Teller is almost completely lost. Experimentally, the lack of the Jahn-Teller distortion is observed\cite{Sanchez06} for x$>$0.6. The increase of the bond angles is more evident in the plane without Ga atoms, since the Mn-O-Mn bond angle is greater than the Ga-O-Mn bond angle.
Since from the results of Table \ref{GaOMnangle}, the Ga-doping just slightly reduces the octahedral distortions with respect to the undoped case, we will concentrate on the JT effect rather than the bond angles.

\begin{table}[!ht]
\caption[{[tabhop}]{Geometric angles as a function of the doping in the plane $ab$. There are empty spaces in the Table because these values are not allowed by the geometrical arrangement of the atoms in the  used supercells.
}
\label{GaOMnangle}
\begin{center}
\begin{tabular}{|l|l|l|}
\hline
Doping & Mn-O-Mn angle & Ga-O-Mn angle \\ 
\hline
x=0.000 & 160$^\circ$ &     \\ 
\hline
x=0.125 & 163$^\circ$ & 162$^\circ$ \\ 
\hline
x=0.250 & 164$^\circ$ & 162$^\circ$ \\ 
\hline
x=0.500 &     & 164$^\circ$ \\ 
\hline
\end{tabular}
\end{center}
\end{table}

We calculate, in Table \ref{DistanceOct} and \ref{DistanceOctNeighbor}, the Mn-O and the Ga-O distances for all the kinds of octahedra: the gallium octahedron, the octahedron of the Mn first-neighbour in the $ab$ plane of the Ga atom, the octahedron of the Mn first-neighbour along the $c$ direction of the Ga atom and the MnO$_6$ octahedra far from the gallium. We find a consistent reduction of the Jahn-Teller effect, the long bonds are shorter than the undoped case and the short bonds become longer. The JT reduction exhibits a robust but not homogeneous trend since it depends on the distance of the Mn-octahedra from the Ga impurities. For instance, the first-neighbour in the $ab$ plane at $x$=0.250 has Mn-O bonds similarly to the mother compound, while all the other octahedra show weaker JT effect.

\begin{table}[!ht]
\caption[{[tabhop}]{Geometric distances of the GaO$_6$ and MnO$_6$ octahedra that are not first-neighbours of gallium. We call them long (l), short (s) and medium (m). The unit is angstrom. There are empty spaces in the Table because these values are not allowed by the geometrical arrangement of the atoms in the used supercells.
}
\label{DistanceOct}
\begin{center}
\begin{tabular}{|l|l|l|l|l|l|l|}
\hline
Doping & Mn-l & Mn-s & Mn-m & Ga-l & Ga-s & Ga-m \\
\hline
x=0.000 & 2.120 & 1.949 & 1.989 &  &   &    \\
\hline
x=0.125 & 2.10  & 1.95  & 1.97  &  2.061 & 1.949  & 2.007   \\
\hline
x=0.250 & 2.094 & 1.947 & 2.000 &  2.088 & 1.924  & 2.021  \\
\hline
x=0.500 &       &       &       &  2.053 & 1.947  & 2.005  \\
\hline
\end{tabular}
\end{center}
\end{table}

\begin{table}[!ht]
\caption[{[tabhop}]{Geometric distances of the MnO$_6$ octahedra first-neighbours of GaO$_6$ octahedra. Mn $ab$ is the manganese atom first-neighbour of the Ga atom in the $ab$ plane, instead Mn $c$ is the manganese atom first-neighbour along the $c$-axis. We call them long (l), short (s) and medium (m). The unit is angstrom. There are empty spaces in the Table, because these values are not allowed by the geometrical arrangement of the atoms in the used supercells.
}
\label{DistanceOctNeighbor}
\scriptsize
\begin{center}
\begin{tabular}{|l|l|l|l|l|l|l|}
\hline
Doping & Mn ab-l & Mn ab-s & Mn ab-m & Mn c-l & Mn c-s& Mn c-m\\
\hline
x=0.000 &  &  &  &  &   &    \\
\hline
x=0.125 & 2.119/2.088  & 1.945/1.975  & 2.005/1.975  & 2.095 & 1.950  & 1.965   \\
\hline
x=0.250 & 2.137 & 1.960 & 1.977 & 2.093 & 1.948  & 1.965  \\
\hline
x=0.500 & 2.067 & 1.952 & 2.002 &  &   &   \\
\hline
\end{tabular}
\end{center}
\end{table}
\normalsize

\subsection{Density of states}

\begin{figure}[t]
\centering
\includegraphics[width=0.35\textwidth, angle=270]{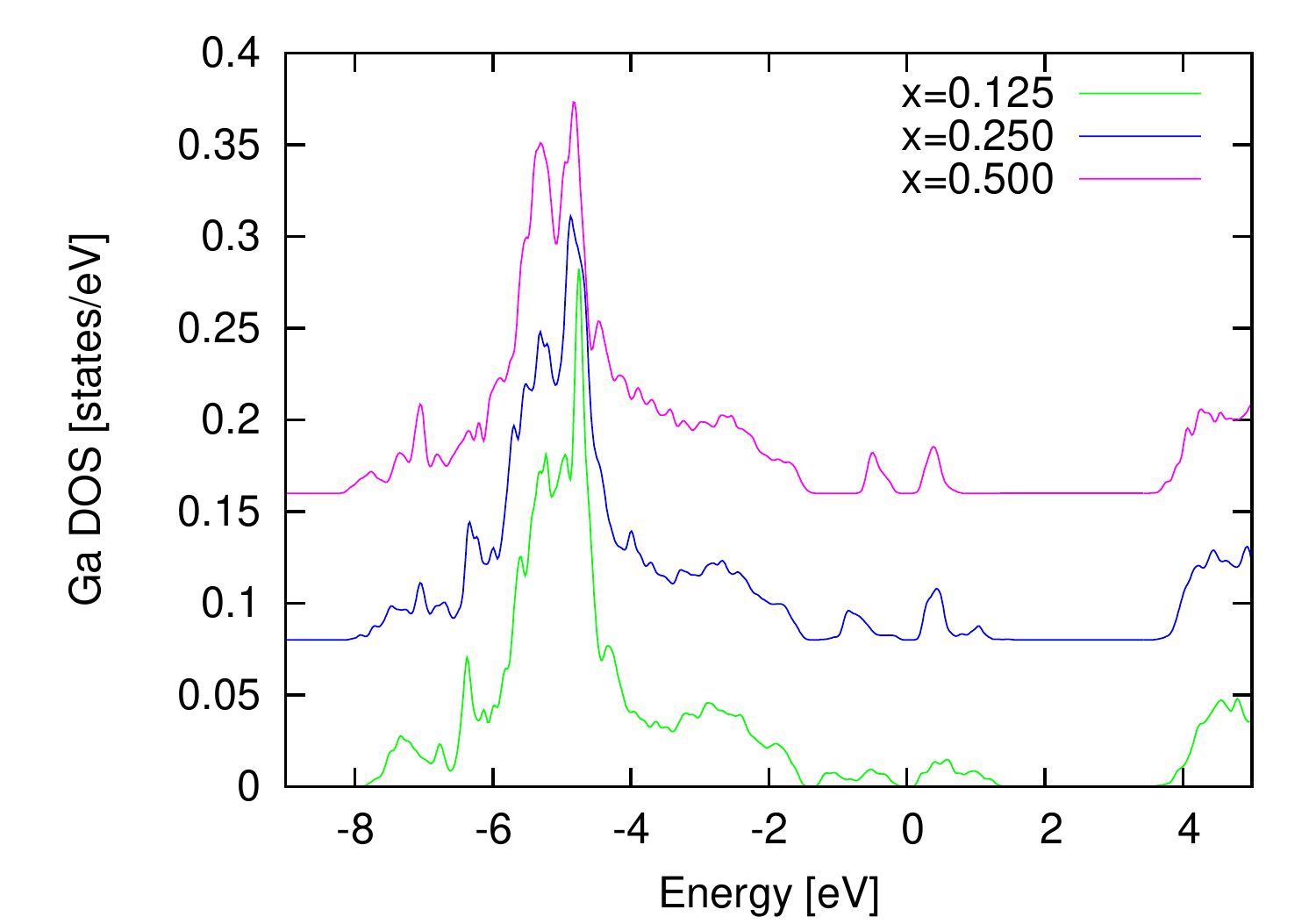}
\includegraphics[width=0.35\textwidth, angle=270]{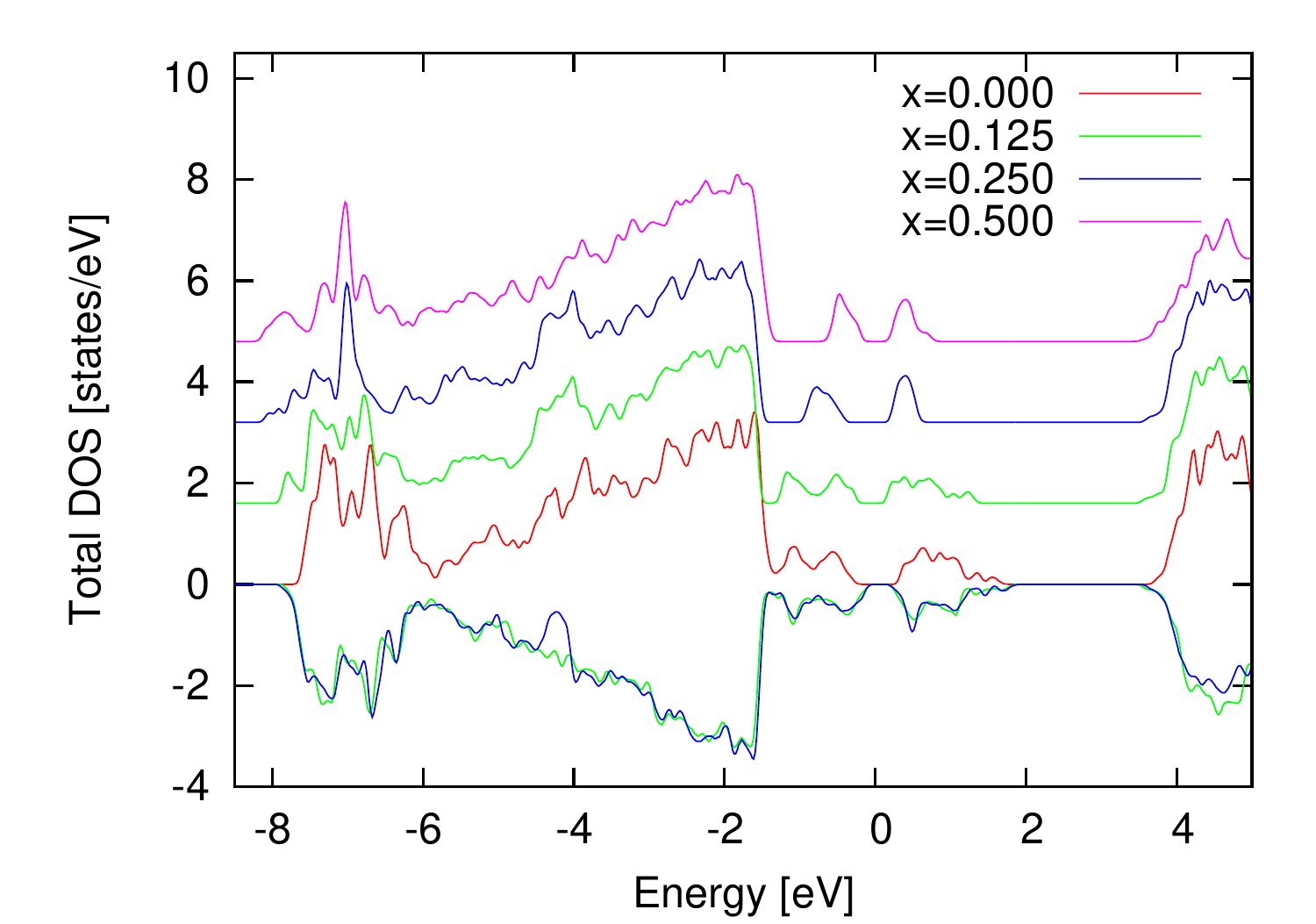}
\caption{Density of state of gallium atom (upper panel) and total density of state (lower panel) per formula unit.
The low energy gallium contribution is negligible.  The DOS contribution between -6~eV and -2~eV is due to the oxygens, while  from -8~eV to -6~eV is due to the $t_{2g}$ of the majority spin.}
\label{fig:DOS}
\end{figure}

To better investigate the effects of the reduced Jahn-Teller distortion, let us now calculate the density of state of our system. To this end, we put the Ga atom in the spin-up plane so that we break the equivalence between the spin-up and spin-down properties, having spin-up and spin-down DOS contributions different also in the AFM phase. From Fig. \ref{fig:DOS}, upper panel, we see that the Ga doping does not affect the low energy part of the DOS since in this regime the Ga ions behave like vacancies. We point out that a non-vanishing contribution appears as a consequence of the small hybridization with the manganese atoms.  This is observed in the metallic system La$_{\frac{2}{3}}$Sr$_{\frac{1}{3}}$MnO$_3$ too, where the Ga-doping removes electronic states at the Fermi level creating an insulating compound \cite{Nucara08}. Moreover, due to the Coulomb repulsion and the intra-orbital $t_{2g}$ and the $e_g$ oxygen contributions, the system behaves like a charge-transfer insulator. Finally, we note that the Jahn-Teller distortion decreases as x increases. Thus, since part of the gap is due to the JT distortions, increasing the gallium doping gives rise to a reduction of the JT distortion lowering the gap.

\begin{figure}[t]
\centering
\includegraphics[width=0.46\textwidth]{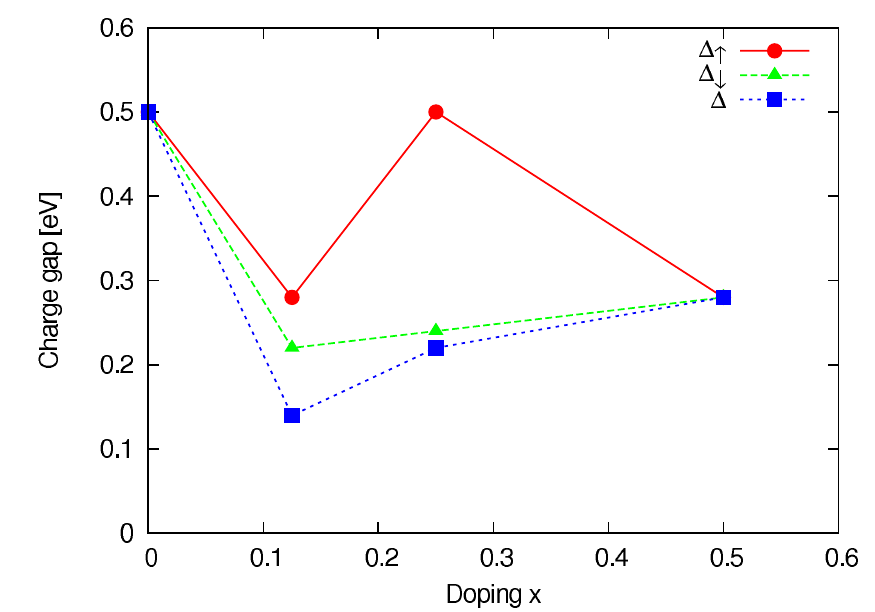}
\caption{Evolution of the gap as a function of doping. We find a strong reduction of the gap at $x$=0.125
due to the breaking of the cooperative Jahn-Teller effect. At $x$=0.250 the gap increases again because the number of magnetic first neighbor decreases. The spin channel $\uparrow$ has a greater gap in comparison
to the spin channel $\downarrow$, because the Ga atom in the spin channel $\uparrow$ reduces the number of magnetic first neighbor.}
\label{fig:GAP}
\end{figure}

However, when we increase the Ga concentration at x=0.250, once more, the gap increases again because the
diminution of the number of magnetic first neighbors as we can see in Fig. \ref{fig:GAP}.
At $x$=0.125, there is a gap only because of the Coulomb repulsion.
We also find that the in-plane gap where the Ga is located coincides with the gap of spin-up channel ($\Delta_{\uparrow}$), whereas the gap in the other plane matches the spin-down channel gap ($\Delta_{\downarrow}$). Moreover, the spin-up gap is always greater than the spin-down one since the Ga doping reduces the number of magnetic first neighbors. This reduction also affects the bandwidth of e$_g$ electrons strongly suppressing it near the gap for the electrons in Ga plane. On the contrary, for down-spin channel the bandwidth of $e_g$ electrons increases up to $x$=0.25 doping, as one can infer looking at Table V. Finally, when $x$=0.5, we find a shrinking of the bandwidth in both spin channels because at this doping both channels contain a Ga atom that, as previously noted, reduces the number of magnetic first neighbors.

\begin{table}[!ht]
\caption[{[tabhop}]{Bandwidths of the $e_g$ electrons for spin-up (W$_\uparrow$) and spin-down (W$_\downarrow$) as a function of the Ga-doping. The unit is eV. The plane with spin-up Mn contains the Ga atom at $x$=0.125 and $x$=0.250. Instead, at $x$=0.500 both planes contain a Ga atom in order to keep the AFM phase.
}
\label{Bandwidth}
\begin{center}
\begin{tabular}{|l|l|l|}
\hline
Doping & $W_{\uparrow}$ & $W_{\downarrow}$ \\
\hline
x=0.000 & 1.45 & 1.45   \\
(experimental atomic positions) &  &  \\
\hline
x=0.000 & 1.48 & 1.48    \\
(relaxed position at exp. volume) & & \\
\hline
x=0.125 & 1.32 & 1.68 \\
\hline
x=0.250 & 0.52 & 1.70 \\
\hline
x=0.500 & 0.66 & 0.66 \\
\hline
\end{tabular}
\end{center}
\end{table}

\subsection{Orbital order}

To investigate the orbital order, we use the following notation \cite{Autieri2014KCrF3}
\begin{equation} 
\label{eq:theta_Mn}
 \mid \theta > =\cos{\frac{\theta}{2}} |3z^2-r^2\rangle{} + \sin{\frac{\theta}{2}} |x^{2}-y^{2}\rangle{}
\end{equation}
where $|3z^2-r^2\rangle{}$ and $|x^{2}-y^{2}\rangle{}$ represent the $e_g$ eigenstates of manganese. We use the 5$\times$5 occupation number matrix to determine $\mid \theta >$ by following the formula (\ref{eq:theta_Mn}) and calculate the $\theta$ angle in the local reference system for all the octahedra for the investigated concentrations.
From Eq. \ref{eq:theta_Mn}, we get that for $\theta$=90$^\circ$, we have 50\% occupation for each orbital, while deviation from this value produces the orbital order that is also directly correlated to Jahn-Teller effect. Besides, for $\theta$=$0^\circ$ we have $\mid \theta >$= $|3z^2-r^2\rangle{}$, while for $\theta$=$180^\circ$ we have $\mid \theta >$ = $|x^{2}-y^{2}\rangle{}$.
For $\theta\approx$143$^\circ$ (37$^\circ$), we obtain that $\tan\frac{\theta}{2}=3(\frac{1}{3}$). These values are of special interest since we have $ \mid \theta >$=$|3x^2-r^2\rangle{}$ or $|3y^2-r^2\rangle{}$. In case, x or y represent the long bond we would have that $ \mid \theta >$=$|3l^2-r^2\rangle{}$.\\

\begin{figure}[t!]
\begin{center}
\includegraphics[width=0.51\textwidth]{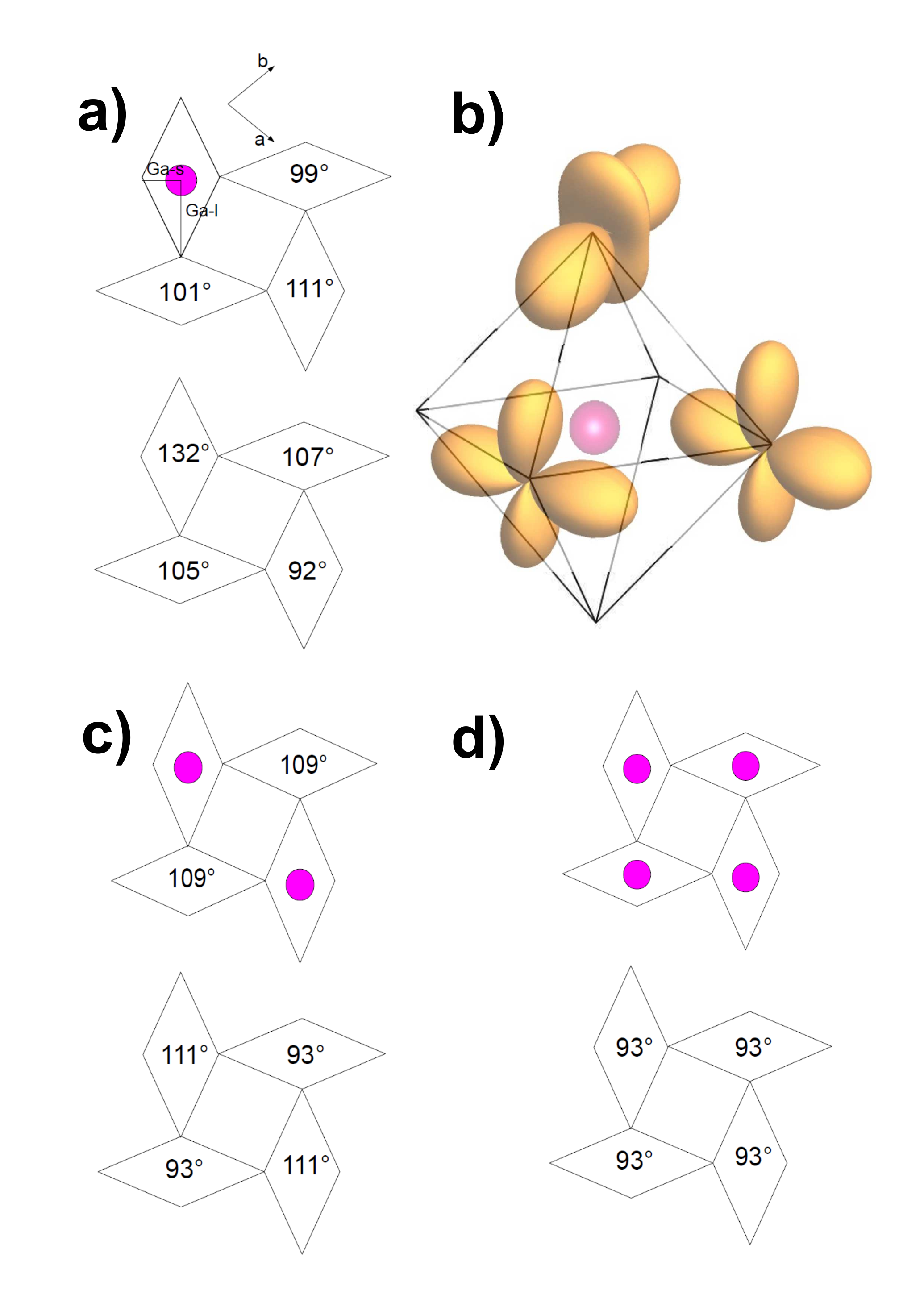}
\caption{
(a) Orbital order for the Mn atoms at $x$=0.125. The $3z^2-r^2$ orbital up the gallium octahedron is strongly suppressed, because does not hybridize with the gallium atom. Excluding the latter octahedron, a $G$-type orbital order structure is observed.
(b) Orbital order of the Mn-atoms for the in-plane and out-of-plane neighbors of the Ga at x=0.125. The in-plane neighbours are almost similar since they present $\theta$=99$^{\circ}$ and 101$^{\circ}$, while the out-of-plane neighbour has $\theta$=132$^{\circ}$
(c) Orbital order for the Mn atoms at $x$=0.250. 
(d) Orbital order for the Mn atoms at $x$=0.500. 
The Ga atom is pink. Ga-s and Ga-l are the short and the long bond of the gallium octahedron.}
\label{fig:Cartoon500}
\end{center}
\end{figure}

For the undoped case, we find a theoretical value of $\theta$=103$^{\circ}$ while the experimental orbital order\cite{Rodr98} is
$\theta$=108$^{\circ}$.
The numerical results for $x$=0.125 and $x$=0.250 are plotted in Fig. \ref{fig:Cartoon500} (a) and (c), respectively. If we consider the difference between the undoped value of $\theta$=103$^{\circ}$ and the considered cases, this has a tendency to a $G$-type structure, that consists in alternated orbital directions. This $G$-type orbital order is that predicted for the hypothetical insulating ferromagnetic phase in LaMnO$_3$ \cite{Hotta99}. Thus, we argue that the gallium doping may induce an instability towards a new kind of orbital order that can drive the ferromagnetism in LaMn$_{1-x}$Ga$_{x}$O$_3$. This G-type orbital order has been predicted for low values of the interaction between the spins of $t_{2g}$ electrons $J'$  \cite{Hotta99},  even though in our case the reduction of the number of magnetic first neighbors might reduce the effective value $J'$. Moreover, this $G$-type orbital order is in agreement with the octahedral distortions discussed in the previous Section, indicating that stronger is the Jahn-Teller higher is $\theta$. The only octahedron that is not in agreement with the $G$-type orbital order is the Mn above the Ga atom at $x$=0.125. In this case the $|3z^2-r^2\rangle{}$ is strongly suppressed leaving room for the creation of an orbital close to $|3l^2-r^2\rangle{}$. The orbitals for the first neighbour of the Ga at $x$=0.125 are represented in Fig. \ref{fig:Cartoon500} (b), the orbital of the first NN along the c-axis has a $|3l^2-r^2\rangle{}$ character, while the reduction of $\theta$ for the in-plane first NN produces a reduction of the orbital order.

At $x$= 0.500, the $\theta$ value for the two equivalent Mn ions is $93^\circ$ and the orbital order is strongly suppressed as shown in Fig. \ref{fig:Cartoon500}(d).
The reduced number of magnetic first neighbors suppresses the cooperative Jahn-Teller effect and strongly modifies the orbital order. This behaviour is different from that exhibited by cubic LaMnO$_3$ since in this last case, the orbital order is still present even though there are no distorsions  \cite{Zenia05}. Therefore, the Ga impurity is more effective than the pressure to reduce and to tune the orbital order.

Reminding the DOS and considering the orbital order, we may state that at $\theta\approx$90$^\circ$ the system is correlated and exhibits a DOS with a reduced band gap. 
These results further suggest that the reduction of octahedral distortions, and subsequently of the $\theta$ value, tends to close the gap. However, a bandwidth reduction due to the absence of Mn states avoids the vanishing of the band gap.

\section{Magnetism}

Experimentally, a large increase in the magnetization as a function of the doping is found in applied magnetic field\cite{Ver02} at low Ga-concentration, while the ferromagnetism is found at intermediate concentration x=0.5-0.6, whereas superparamagnetism appears at x=0.8-0.9\cite{Blasco}.
We will investigate separately the two first cases.
The {\it ab-initio} magnetic ground state of LaMnO$_3$ is a non straightforward problem\cite{Sawada97} and the ferromagnetic phase is always metallic, therefore it is difficult to obtain the insulating ferromagnetic ground state. 
To this end, we calculate the energy difference between the antiferromagnetic phase and the ferromagnetic phase for all Ga concentrations,
finding clear indications that the antiferromagnetic phase becomes weaker increasing the gallium doping.

\subsection{Spin-flip at low concentration of Ga}

At the low concentration of 12.5\% of Ga-doping, the isolated gallium atom is surrounded by manganese atoms in the B-site sublattice, where for B-site we define the transition metal site of the ABO$_3$ perovskite. Concerning the magnetic moment, it has been suggested that the magnetic moment in an external applied magnetic field could be due to a spin-flip of manganese atoms first-neighbor to the impurity along the c-axis\cite{Farrell04}.  However, we find that this spin-flip is strongly energetically disfavored in our first-principle calculations. Instead, we actually find that the favored spin-flip is related to the Mn first-neighbor to the Ga atoms in the ab-plane. Therefore, we argue that at low concentration of Ga-doping and external magnetic field a spin-flip can happen within the plane of the gallium dopant. This picture can be easily understood within the Ising approximation of the Mn localised moments. Indeed, the in-plane ferromagnetic coupling\cite{Solovyev96} in LaMnO$_3$ is hard to break when there are four first-neighbors. However, the Ga dopant reduces the number of magnetic first-neighbors implying that the Mn magnetic moments may be easier orientated in the applied magnetic field.

\subsection{Cationic order at intermediate Ga-concentration}

\begin{figure}[t]
\begin{center}
\hbox{
\includegraphics[width=0.25\textwidth]{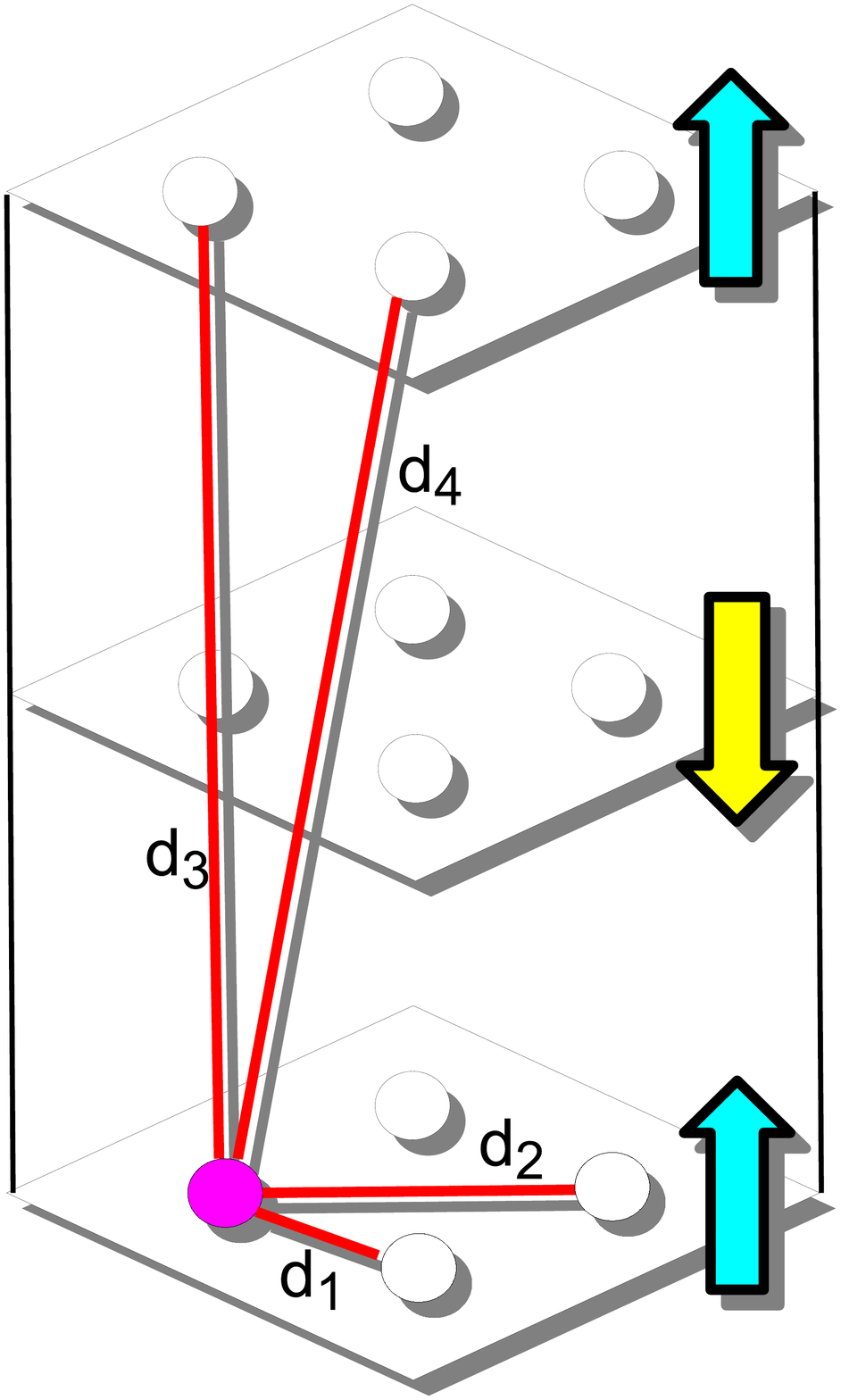}
\includegraphics[width=0.25\textwidth]{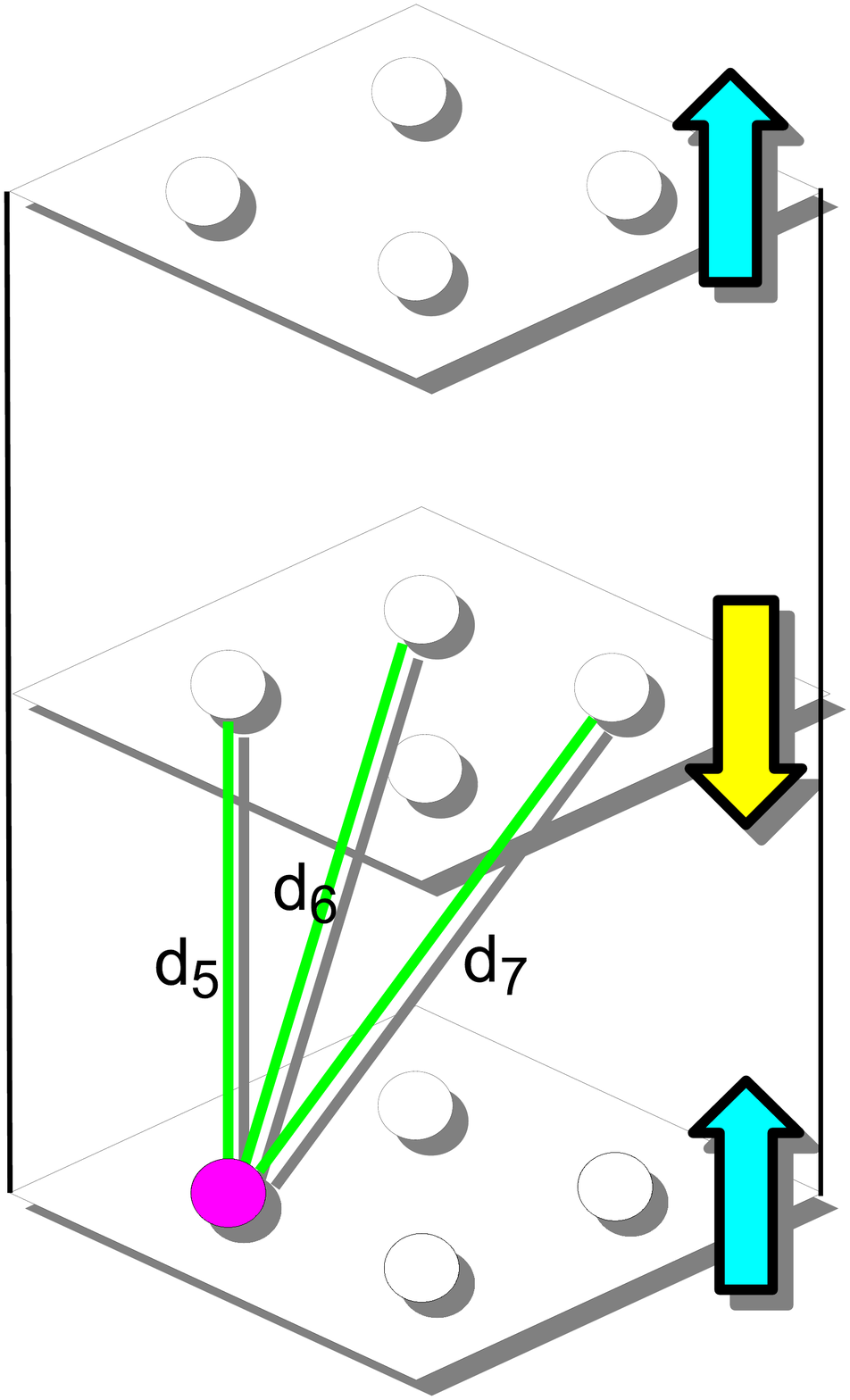}
}
\includegraphics[width=0.40\textwidth]{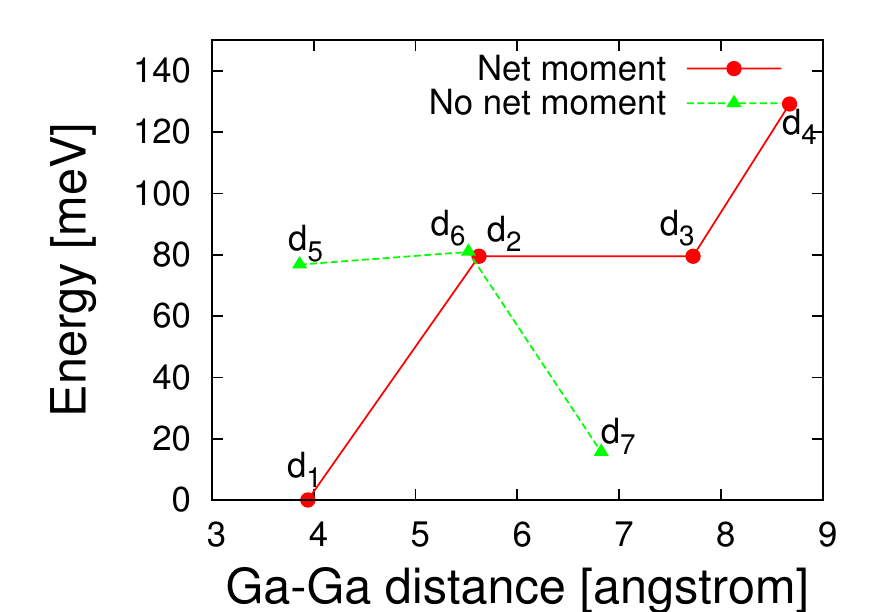} 
\caption{In the top panel, we show the non equivalent space configurations for the two Ga atoms in the supercells at $x$=0.250. The magnetic atoms in different planes have different spin orientations due to the A-type magnetic ordering, the blue arrows represent the spin-up plane while the yellow arrows represent the spin-down plane. In the left top panel, the red lines link two Ga atoms located at layers with the same spin channel. 
In the right top panel, the green lines link two Ga atoms located at layers with different spin channels.
In the bottom panel, we show the energy differences as a function of the Ga-Ga distance in the supercells with 8 octahedra at $x$=0.250.
We plot the energy differences per Ga atom. The red points represent the magnetic configurations with a net magnetic moment, while the green points represent the magnetic configuration with no net magnetic moment. $d_{i}$ with i=1,..,7 are the distances defined in the top panel. The lines are guides for the eyes. The ground state is the magnetic configuration when two Ga atoms are first-neighbours in the $ab$ plane.}
\label{fig:Notation}
\end{center}
\end{figure}

At the intermediate concentration of 25\% of Ga-doping, a long-range ferromagnetic phase it is experimentally found.
We analyse the possible occurrence of a correlated disordered configuration for the Ga-doping by inserting two Ga atoms at different distances in the supercell at $x$=0.250 in the AFM phase.
In the AFM configuration, two types of doping configurations are possible: configurations with and without net magnetic moment. In the first case, two Ga atoms substitute two Mn atoms with the same spin, creating a net magnetic moment and non-compensated ferromagnetism. In the second case, the two Ga atoms substitute two Mn atoms with different spin leaving the net magnetization equal to zero. We can observe all the possible configurations with supercells of 8 octahedra in the top panel of Fig. \ref{fig:Notation}.
In the bottom panel of Fig. \ref{fig:Notation}, we set to zero the ground state energy and plot the energy of the system per formula unit as a function of the Ga-Ga distance in the supercell. 
We find that the ground state is achieved when two Ga atoms are at distance $d_1$, therefore when they are first-neighbour in the $ab$ plane. 
The nonmagnetic Ga-atoms arrange in a cationic order that influences also the magnetic properties of the compound. 
This produced a layered ordered double perovskite, alternated Ga and Mn layers giving rise to a ground state with a net magnetic moment.
This result is easy to understand if we consider that, moving the only planar oxygen between the two Ga atoms, it is possible to reduce the distortion of the GaO$_6$ octahedra without modifying the MnO$_6$ octahedra. In this way, the system minimizes the elastic energy and the total energy, hence creating a ferromagnetic ground state. For completeness, we notice that the other configurations are higher in energy. In particular, when the Ga-Ga distance is $d_5$ we have the two Ga atoms first-neighbour along the $c$-axis. Nevertheless,  the dynamics of the apical oxygen can not produce any elastic energy gain since the distortions lie in the $ab$ plane. Indeed, the relevant structural change between the real orthorhombic structure of LaMnO$_3$ and a hypothetical cubic structure is related to the position of the planar oxygens.

Thus, we propose that the Ga-doping creates layered ordered double perovskite exhibiting a non-compensated ferromagnetism. This is an example of cationic order observed in perovskites. This effect, together with a weaker AFM phase and the instability towards a $G$-type structure of the orbital order, can destroy the AFM phase favoring the onset of a long-rang ferromagnetic order. We note that the non-compensated ferromagnetism alone is not sufficient to explain the ferromagnetic moment experimentally found at
intermediate doping \cite{Blasco}, that has a large magnetic moment per Mn atoms, but
helps the system to cross to the ferromagnetic insulating phase. 

\subsection{Ferromagnetism in layered order LaMn$_{0.5}$Ga$_{0.5}$O$_3$}

Near $x$=0.500 the system exhibits an insulating ferromagnetic phase which is quite difficult to reproduce by using {\it ab-initio} approaches.
Differently from the metallic ferromagnetic phase, the non-compensated ferromagnetism is one possibility to reproduce
a ferromagnetic insulator. 
This insulating ferromagnetic phase is also different from the cubic phase obtained under pressure by orbital splitting. Indeed, up to a pressure of 32 GPa, the system is a paramagnetic metal at room temperature and an antiferromagnetic metal at low temperatures, though this region of the phase diagram is not well explored\cite{Pastorino11}.

\begin{figure}[t]
\begin{center}
\includegraphics[width=0.45\textwidth]{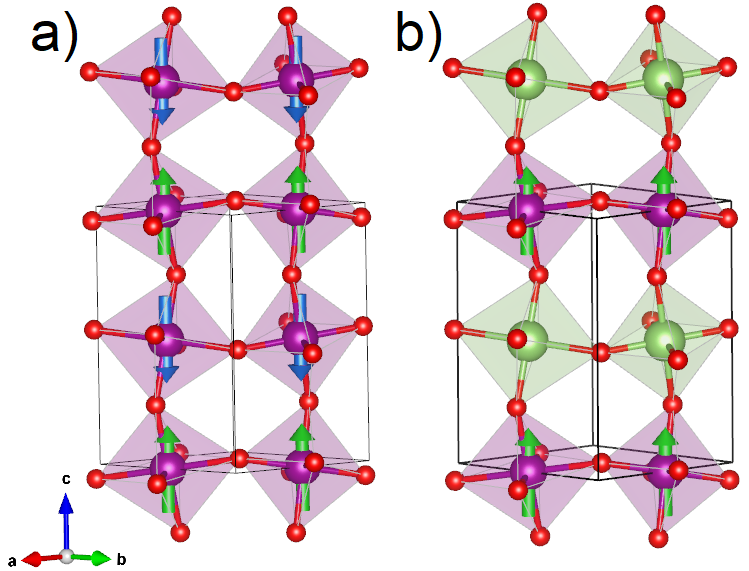}
\caption{a) Undoped LaMnO$_3$ with A-type antiferromagnetic order. b) Layered order arrangement of the transition-metal site in LaMn$_{0.5}$Ga$_{0.5}$O$_3$. The Ga atoms replace the spin-down Mn atoms. MnO$_6$ octahedra are shown in purple while GaO$_6$ octahedra are shown in light green. Red balls represent the oxygen atoms, while the La atoms are not shown. Green and blue arrows represent spin-up and spin-down, respectively.}
\label{fig:layered}
\end{center}
\end{figure}

At $x$=0.500, the ground state is composed by layers of Mn alternated with layers of Ga as shown in Fig. \ref{fig:layered}. 
This layered order arrangement of the B-site with ferromagnetic layers is quite unusual among perovskites. 
Another layered ordered ferromagnetic perovskite is LaCu$_{0.5}$Sn$_{0.5}$O$_3$\cite{layereddouble1,layereddouble2}.
Analogously to LaMn$_{1-x}$Ga$_{x}$O$_3$, this perovskite has La as A-site, large octahedra with magnetism and strong JT together with small octahedra with d$^{10}$ electronic configuration. This is one of the few known examples of layered ordered double perovskite\cite{layereddouble3}.
However, many cases of alternate layers of different phases are known in transition metal perovskites\cite{Autieri12,Autieri14SRO}.\\

\section{Conclusions}

We have calculated the density of state, the evolution of the orbital order and the octahedral distortions for LaMn$_{1-x}$Ga$_{x}$O$_3$. 
The Ga-doping produces an effective orbital and electronic vacancy without creating any hole-doping.
We find that the gallium doping, which slightly reduces octahedral distortion, energetically disfavors the A-type antiferromagnetic phase. The Ga-doping favors the G-type orbital order and weakens the Jahn-Teller effect highlighting electronic correlations.
We argue that the large magnetization at low concentration in an applied magnetic field can be explained by a spin-flip of the in-plane first-neighbours of the gallium. We find that the ferromagnetic phase at x=0.5 is favored by a ferromagnetic layered ordered double perovskite.

Additional mechanisms for the creation of large T$_C$ ferromagnets typical of the DMS could be present, however, their investigation is beyond the task of this paper.

\medskip

\begin{acknowledgments}
We acknowledge useful discussions with M. Le\v{z}ai\'{c}.
The work is supported by the Foundation for Polish Science through the International
Research Agendas program co-financed by the European
Union within the Smart Growth Operational Programme.
\end{acknowledgments}


\end{document}